\documentclass[10pt,twocolumn,prl,english,showpacs]{revtex4}
\usepackage[T1]{fontenc}
\usepackage[latin9]{inputenc}
\usepackage{float}
\usepackage{graphicx}
\usepackage{amssymb}
\usepackage{amsmath}
\RequirePackage{xspace}

\makeatletter

\usepackage{color}
\usepackage{graphicx}
\usepackage{booktabs}
\usepackage{array}
\usepackage{calc}
\usepackage{dcolumn}

\newcommand{\ie}{i.\,e.\@\xspace}

\newcommand{\Eq}{Eq.\@\xspace}

\newcommand{\shalf}{{\textstyle \frac{1}{2}}}

\DeclareMathOperator*{\Arg}{Arg\!}

\usepackage{babel}
\makeatother

\begin{document}

\title{Synchronizing distant nodes: a universal classification of networks}

\author{V. Flunkert\textsuperscript{1}}
\author{S. Yanchuk\textsuperscript{2}}
\author{T. Dahms\textsuperscript{1}}
\author{E. Sch\"oll\textsuperscript{1}}
\email{schoell@physik.tu-berlin.de}

\affiliation{\textsuperscript{1}Institut f{\"u}r Theoretische Physik,
  TU Berlin, Hardenbergstra\ss{}e 36, 10623 Berlin, Germany}

\affiliation{\textsuperscript{2}Institut f{\"u}r Mathematik, Humboldt
  Universit{\"a}t Berlin, Unter den Linden 6, 10099 Berlin, Germany}

\begin{abstract}
  Stability of synchronization in delay-coupled networks of identical
  units generally depends in a complicated way on the coupling
  topology.  We show that for large coupling delays synchronizability
  relates in a simple way to the spectral properties of the network
  topology.  The master stability function used to determine stability
  of synchronous solutions has a universal structure in the limit of
  large delay: it is rotationally symmetric around the origin and
  increases monotonically with the radius in the complex plane.  This
  allows a universal classification of networks with respect to their
  synchronization properties and solves the problem of complete
  synchronization in networks with strongly delayed coupling.
\end{abstract}


\maketitle

Synchronization phenomena in networks are of great importance
\cite{PIK01} in many areas.  Chaos synchronization of lasers, for
instance, may lead to new secure communication schemes
\cite{CUO93BOC02KAN08a}. The synchronization of neurons is believed to
play a crucial role in the brain under normal conditions, for instance
in the context of cognition and learning \cite{SIN99}, and under
pathological conditions such as Parkinson's disease \cite{TAS98}.
Time delay effects are a key issue in realistic networks. For example,
the finite propagation time of light between coupled semiconductor
lasers \cite{WUE05aCAR06ERZ06aFIS06DHU08} significantly influences
the dynamics. Similar effects occur in neuronal \cite{ROS05MAS08} and
biological \cite{TAK01} networks.

To determine the stability of a synchronized state in a network of
identical units, a powerful method has been developed
\cite{PEC98}, i.e., the master stability function (MSF). Recent
works \cite{DHA04CHO09,KIN09} have started to investigate the MSF for
networks with coupling delays and found that the MSF depends
non-trivially on delay times.

In this work we show that in the limit of large coupling delays the
MSF has a very simple structure. This solves the problem of complete
zero-lag synchronization for networks with large coupling delay. After
briefly introducing the notion of the MSF, we demonstrate the
implications for large coupling delays based on a scaling theory
\cite{FAR82GIA96MEN98WOL06YAN06YAN09}.  This allows us to describe the
synchronizability of networks with strongly delayed coupling depending
on the type of node dynamics and spectral properties of the network
topology. For example, as recently conjectured \cite{KIN09}, networks
for which the trajectory of an uncoupled unit is also a solution of
the network cannot exhibit chaos synchronization for large coupling
delay.  The results presented here confirm and generalize these
previous findings.

Consider a system of $N$ identical units connected in a network with a
coupling delay $\tau$ \cite{KIN09} ($x^i\in\mathbb{R}^d$,
$i=1,\dots,N$)
\begin{equation} 
  \dot  x^i (t) =  f \left[ x^i(t)\right] + \sum\nolimits_{j=1}^N g_{ij}  h\left[ x^j(t-\tau)\right]. \label{eq:network}
\end{equation}
Here, $g_{ij}$ is the real-valued coupling matrix, which determines
the topology and the strength of each link in the network, $f$ is a
(non-linear) function describing the dynamics of an isolated unit, and
$h$ is a possibly non-linear coupling function. To allow for an
invariant synchronization manifold (SM), the row sum $\sigma =
\sum_{j=1}^N g_{ij}$ of the matrix has to be the same for each row $i$
\cite{PEC98}.  The stability of the synchronized solution is then
governed by the MSF and the eigenvalues of the coupling matrix
$g_{ij}$. The MSF is defined as the maximum Lyapunov exponent
$\lambda_{\rm max}(re^{i\psi})$ as a function of the complex argument
$r e^{i\psi}$ arising from the variational equation
\begin{equation*} 
  \dot \xi(t) = D f[x(t)]\, \xi(t) + r e^{i\psi} Dh[x(t-\tau)]\,\xi(t-\tau),
\end{equation*}
where $x(t)$ is given by the dynamics within the SM. The synchronized
state is stable for a given coupling topology if the MSF is negative
at all transversal eigenvalues $\gamma_k$ of the coupling matrix
($\lambda_{\rm max}(\gamma_k)<0$).  Here, transversal eigenvalue
refers to all eigenvalues except for the eigenvalue $\sigma$
associated to perturbations within the SM with corresponding
eigenvector $(1,\,1,\,\dots,\,1)$.

We will now restrict our analysis to maps \cite{KIN09},
but all ingredients of our argument are also valid for flows. For
delay-coupled maps the dynamics in the SM is governed by the
equation $x_{k+1} = f(x_k) +  \sigma h(x_{k-\tau})$ with $\tau\in\mathbb{N}$ and $x_k\in\mathbb{C}^d$ or $\in\mathbb{R}^d$
and the MSF is calculated for fixed $\sigma$ from
\begin{equation} 
  \xi_{k+1} =  A_k \xi_k + r e^{i\psi}  B_k \xi_{k-\tau} \label{eq:LD-var}
\end{equation}
with matrices $A_k=Df(x_k)$  and $B_k=Dh(x_{k-\tau})$.

Note that when the delay is changed the dynamics in the SM changes,
too.  Hence, we are not able to make predictions about what happens as
$\tau$ is changed. However, at a fixed large value of the delay time
$\tau$ we can analyze the Lyapunov exponents arising from different
values of $re^{i\psi}$ in \Eq \eqref{eq:LD-var}. We do this in the
following steps: first we analyze the two simpler cases when the
dynamics in the SM is a fixed point (FP) or a periodic orbit
(PO). Then, to expand the results to chaotic dynamics in the SM, we
use the fact that POs are dense in a chaotic attractor.

For FPs and POs of delay differential equations a scaling theory for
the eigenvalues or Floquet exponents in the limit of large delay
\cite{FAR82GIA96MEN98WOL06YAN06YAN09} shows that the spectrum
consists in both cases of two parts: a strongly unstable part arising
from unstable eigenvalues of the system without delay and a
pseudo-continuous spectrum for which the real parts of the eigenvalues
approach zero in the limit of large delay. This scaling theory has
been developed for flows; to prove our statements we will extend this
theory to maps.

\textsl{Fixed point} -- Let us first consider the case of a FP in the
SM, for which  $A=A_k$ and $B=B_k$ are constant.  Making the ansatz $\xi_k =
z^k \xi_0$, we find an equation for the multipliers~$z$
\begin{equation} 
  \det[A -z I + re^{i\psi} B\, z^{-\tau}] = 0, \label{eq:char}
\end{equation}
where $I$ denotes the identity matrix.

For the strongly unstable spectrum we suppose there is a solution with
$|z|>1$. Then in the limit of $\tau\to \infty$ \Eq \eqref{eq:char}
becomes $\det[A -z I] = 0$.  Thus in the limit of large delay the
eigenvalues $z$ of $A$ with $|z|>1$ are also solutions of \Eq
\eqref{eq:char} and vice versa.

We are now interested in the pseudo-continuous spectrum, \ie, the
solutions with $|z|\approx 1$ in the limit of large $\tau$.  We make
the ansatz $z=(1+\delta/\tau) e^{i\omega}$.  In the limit $\tau\to
\infty$ we have $(1+\delta/\tau)^{-\tau} \sim e^{-\delta}$ and
$(1+\delta/\tau) \sim 1$, and \Eq \eqref{eq:char} becomes
\begin{equation} 
  \det[A - I e^{i\omega} + r e^{-\delta} e^{i(\psi-\phi)} B ] = 0 \label{eq:LD-charlim}
\end{equation}
with $\phi=\omega\tau$.  As we will show below, $\omega$ as well as
the parameter $\phi$ take on any (arbitrarily dense) values in
$[-\pi,\,\pi]$. From this it is clear that the phase $\psi$ in the
variational equation does not change $\delta$, \ie, the MSF is
invariant under phase shifts (rotations) and its value only depends on
$r$.

Equation \eqref{eq:LD-charlim} is a polynomial in $\mu= r e^{-\delta}
e^{i(\psi-\phi)}$ for which the roots can be calculated. For example,
if $B$ is invertible, the roots $\mu$ are the eigenvalues of the
matrix $-B^{-1}(A-I e^{i\omega})$. In general, each root $\mu$ is a
function of $\omega$ and one can find the branches $\delta(\omega) =
-\ln |\mu(\omega)| + \ln r$ from the definition of $\mu$.  The
function $\mu(\omega)$ can admit the zero value at some point
$\omega_0$, \ie, $\mu(\omega_0)=0$, in the case when the matrix $A$
has an eigenvalue with $|z|=1$. Indeed, as follows from \Eq
\eqref{eq:LD-charlim}, for $\mu=0$, $\omega=\omega_0$ and $\det B\ne
0$ we have $\det [ A- I e^{i\omega_0}] = \det[A-Iz]=0$.  In all other
cases, with $\det B\ne 0$ and $|z|\ne 1$, the function $|\mu(\omega)|$
is bounded $0 < \mu_0 \le |\mu(\omega)| \le \mu_1$.

If there are no strongly unstable eigenvalues, the sign of $\delta$
determines the stability in the limit of large $\tau$, since
$|z| \sim |1+\delta/\tau|$. It is clear that $\delta$ increases
monotonically with increasing $r$ and in particular $\delta$ is
negative for small $r$ and positive for large $r$. Thus there is a
critical radius $r_0$ for which the first eigenvalue branch becomes
unstable ($\delta>0$) and thus the MSF changes sign.

Note that we have obtained the function $\delta(\omega)$ on which the
solutions lie in the limit of large $\tau$ but not yet the exact
values of $\omega$. These values can be calculated from the expression
$\mu(\omega) = r e^{-\delta(\omega)} e^{i(\psi - \omega\tau)}$, which
implies
\begin{equation} 
  \Arg\, \mu(\omega) = \psi -\omega\tau + 2\pi k \label{eq:arg}
\end{equation}
for any integer $k$.  Since $\mu(\omega)$ is a known root of \Eq
\eqref{eq:LD-charlim}, \Eq \eqref{eq:arg} can be considered as a
transcendental equation for determining the solutions
$\omega=\omega_k$.  In particular, \Eq \eqref{eq:arg} implies that the
distance between neighboring solutions $\omega_k$ and $\omega_{k-1}$
\begin{align*} 
  \omega_k-\omega_{k-1} &= [\Arg\, \mu(\omega_{k-1}) - \Arg\,\mu(\omega_k)]/\tau + 2\pi/\tau\\
  &= 2\pi/\tau + \mathcal{O}\left(1/\tau^2\right)
\end{align*}
is proportional to $1/\tau$ and the curve $\delta(\omega)$ is filled
densely with equally spaced roots as $\tau\to\infty$.

\begin{figure}
  \includegraphics[width=\linewidth]{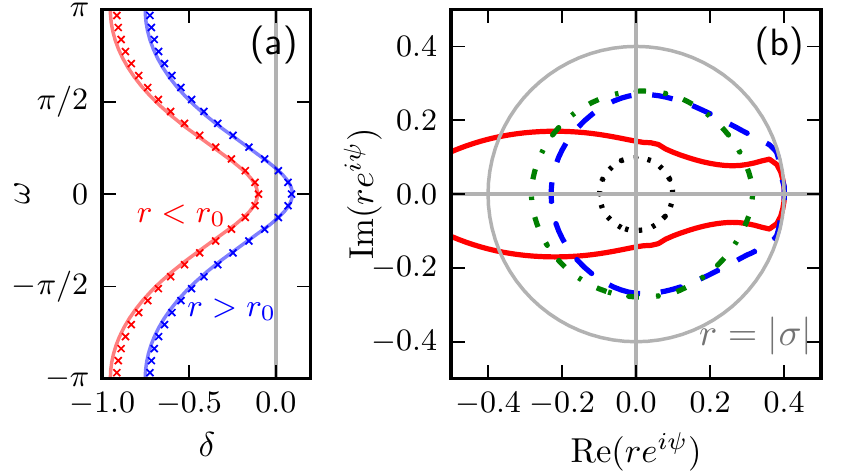}
  \caption{\label{fig:pseudo} (Color online) (a) Pseudo-continuous spectrum
    $\delta(\omega)$ (lines) and location of the exact roots (crosses)
    for a one-dimensional complex map for $r=3.3 > r_0=3$ and $r = 2.7
    < r_0=3$.  Parameters: $A=0.4$, $B=0.2$, $\psi=0$, $\tau=30$.
    (b) Contour line $\lambda_{\rm max}=0$ of the MSF for coupled
    semiconductor lasers according to Eq.~(\ref{eq:lk}) for delay
    times $\tau=1$ (solid), $8$ (dashed), $20$ (dash-dotted), and
    $1000$ (dotted).
}
\end{figure}

For illustration, consider the simple case of a one-dimensional
complex map with $A,\,B\in\mathbb{C}$ with $|A|<1$. In this case we
can explicitly calculate the pseudo-continuous spectrum
$\delta(\omega) = \ln(|r B|/|A-e^{i\omega}|)$, which is depicted in
Fig.~\ref{fig:pseudo}a.  For $r<(1-|A|)/|B|$ all the eigenvalues
approach $|z|=1$ from the stable side and for $r> (1-|A|)/|B|$ there
are always weakly unstable eigenvalues. Thus the critical radius is
given by $r_0=(1-|A|)/|B|$.

\textsl{Periodic orbit} -- Now consider the variational \Eq
\eqref{eq:LD-var} with $A_k$ and $B_k$ being periodic in $k$ with
period $T$, corresponding to a PO in the SM. We consider the case of
large delay, \ie, $\tau \gg T$.  Making a Floquet-like ansatz $\xi_k =
z^k\, q_k$, where $q_k$ is $T$-periodic, we find
\begin{equation} 
  z\, q_{k+1} = A_k\, q_k + re^{i\psi} B_k\, z^{-\tau} q_{k-n } \label{eq:floq}
\end{equation}
with $ n = \tau\, \mathrm{mod} \,T \in \lbrace 0,\,1,\dots,\,
T-1\rbrace$.

For the strongly unstable spectrum again suppose there is a solution
with $|z|>1$, then in the limit $\tau\to\infty$ the term $z^{-\tau}$
vanishes and we find
\begin{equation} 
  z\, q_{k+1} = A_k\,q_k. \label{eq:floq-strong}
\end{equation}
Using the periodicity of $q_k$, \Eq \eqref{eq:floq-strong} implies
$\det [ z^T - \prod_{k=1}^{T} A_k] = 0,$
where $z^T$ is a Floquet multiplier of the system $\xi_{k+1} = A_k
\xi_k$ without delay.  Hence, if $z^T$ is a Floquet multiplier of \Eq
\eqref{eq:floq-strong}, with $|z|>1$, then in the limit
$\tau\to\infty$ it is also a solution of \Eq \eqref{eq:LD-var} and vice
versa.

For the pseudo-continuous spectrum we again make the ansatz
$z=(1+\delta/\tau) e^{i\omega}$. Taking the limit $\tau\to\infty$
\Eq \eqref{eq:floq} becomes
\begin{equation} 
  e^{i\omega}\, q_{k+1} = A_k\, q_k + re^{-\delta} e^{i(\psi-\phi)} B_k\, q_{k - n}\label{eq:1}
\end{equation}
with $\phi=\omega\tau$. Thus one has to solve
\begin{equation} 
  [e^{i\omega} \overline J + \overline A + \mu \overline B]\vec q = 0, \label{eq:floq-pseudo}
\end{equation}
where $\mu=re^{-\delta} e^{i(\psi-\phi)}$ and $\vec q =
(q_1,\dots,\,q_T)$. The matrices $\overline J$, $\overline A$, and
$\overline B$ follow from Eq.~\eqref{eq:1}, taking into account the
periodicity of $A_k$, $B_k$, and $q_k$, e.g., $\overline A =
\mathrm{diag}\lbrace A_1,\dots,\,A_T\rbrace$.  Taking the determinant
of the entire matrix in \Eq \eqref{eq:floq-pseudo} results in a
polynomial in $\mu=re^{-\delta}e^{i(\psi-\phi)}$ (of maximum order
$Td$). Again, the roots $\mu$ are functions of $\omega$ and we can
calculate the branches $\delta(\omega) = -\ln |\mu(\omega)| + \ln r$,
where $\psi$ and $\phi$ drop out. As in the case of FPs, one can show
that the function $|\mu(\omega)|$ is bounded $0<\mu_0\le |\mu(\omega)|
\le \mu_1$ unless the instantaneous system has a Floquet multiplier
$z$ with $|z|=1$.  Note that for the FP case as well as for the PO
case, one can show
that the discussed strongly unstable and pseudo-continuous spectrum
constitute the entire spectrum.

We have found the same structure of the MSF for a PO in the SM: The
MSF is rotationally symmetric about the origin in the complex
plane. If without feedback $(r=0)$ the MSF is positive, then it is a
positive constant in the limit of large delay. Otherwise it is a
monotonically increasing function of $r$ and it changes sign at a
critical radius $r_0$.

\textsl{Chaotic dynamics} -- Every chaotic attractor embeds an
infinite number of unstable periodic orbits (UPOs). It is well known
that the characteristic properties of the chaotic system can be
described in terms of these UPOs.
One of the most important examples is the natural measure of the
chaotic attractor which is concentrated at the UPOs and can in fact be
expressed in terms of the orbit's Floquet multipliers
\cite{GRE88,LAI97}.

Lyapunov exponents arising from variational equations such as \Eq
\eqref{eq:LD-var} have been discussed in the framework of PO theory
\cite{CVI95}, too.  In particular it has been shown \cite{NAG97} that
a chaotic attractor in an invariant manifold loses its transversal
stability in a blow-out bifurcation when the transversely unstable
orbits outweigh the transversely stable orbits. To be precise, we
divide the orbits into these two groups and define \cite{NAG97} the
transversely stable weight $\Lambda_T^{s}$ and the unstable weight
$\Lambda_T^{u}$ as
\begin{align} 
  \Lambda_T^{u,s} &= \sum\nolimits_{j=1}^{N^{u,s}_T} \mu_T(j) \lambda_T(j), \label{eq:weightedsum}
\end{align}
where the sum goes over all $N^u_T$ transversely unstable and $N^s_T$
transversely stable orbits with period $T$ (or factors of $T$),
respectively. Here, $\mu_T(j)$ is the weight of the $j$th orbit,
corresponding to the natural measure of a typical trajectory in the
neighborhood of the $j$th orbit and $\lambda_T(j)$ is the transversal
Lyapunov exponent of this $j$th orbit. The weight of a PO is inversely
proportional to the product of its unstable Floquet multipliers
\cite{GRE88}. The attractor is transversely unstable if and only if in
the limit of large $T$
\begin{equation} 
 \Lambda^u_T  > |\Lambda^s_T|. \label{eq:weightcond}
\end{equation}

We now draw the connection to the scaling theory for large $\tau$.
Starting from $r=0$ (no feedback), the transversal Lyapunov exponents
$\lambda_T(j)$ of each orbit can only increase with increasing $r$, as
shown above, and the weights $\mu_T(j)$ are not changed. In particular
for large enough $r$ any orbit becomes transversely unstable: either
it is already unstable for $r=0$ and thus remains unstable, or the
pseudo-continuous spectrum goes to zero and for large $r$ it does so
from the unstable side. Thus there exists a minimum radius $r_0$ for
which the condition \eqref{eq:weightcond} on the weights is fulfilled.
Note that since we consider the limit $\tau\to\infty$ we can evaluate
\Eq \eqref{eq:weightcond} at arbitrarily large $T$.  Thus in summary
the MSF for chaotic dynamics has the same structure as for FPs and POs
(the rotation symmetry follows from the rotation symmetry of each
$\lambda_T(j)$).

Let us now discuss what the structure of the MSF means for the
synchronizability of networks. We can classify networks into three
types depending on the magnitude of the largest transversal
eigenvalues $\gamma_{\rm max}$ in relation to the magnitude of the row
sum $\sigma$: (a) $|\gamma_{\rm max}| < |\sigma|$, (b) $|\gamma_{\rm
  max}| = |\sigma|$, and (c) $|\gamma_{\rm max}| > |\sigma|$.

As we have shown above, for stable synchronization it is necessary
that $|\gamma_{\rm max}|<r_0$. If $|\gamma_{\rm max}|>r_0$, the
synchronization is not stable. Since $\sigma$ is the eigenvalue of
the coupling matrix associated with the synchronous mode, the MSF
$\lambda_{\rm max}(\sigma)$ describes the local stability within the
SM, i.e., $\lambda_{\rm max}(\sigma)>0$ for chaotic dynamics in the SM
and $\lambda_{\rm max}(\sigma)<0$ for FPs or POs. This implies that
$|\sigma|>r_0$ in the first case and $|\sigma|<r_0$ in the latter
case.  In other words, the row sum $\sigma$ gives an estimate of the
critical radius $r_0$. In particular, it allows us to give a complete
classification (Table \ref{table}).
\begin{table}
  \caption{\label{table} Stability of chaotic and non-chaotic synchronized 
    solutions for the three types of networks}
\renewcommand{\arraystretch}{1.2}
\begin{ruledtabular}
\begin{tabular}{p{2.4cm}|p{3.0cm}|p{4.4cm}}
  & chaotic dynamics\newline in the SM ($r_0< |\sigma|$) & PO or FP in the SM\newline ($|\sigma| < r_0$) \\
  \midrule
  (a) $ |\gamma_{\rm max}|<|\sigma|$ & synchr.\ stable if \newline $ |\gamma_{\rm max}| < r_0 $ & synchr.\ stable \\
  \midrule
  (b) $|\gamma_{\rm max}|=|\sigma|$  & synchr.\ unstable  & synchr.\ stable \\
  \midrule
  (c) $|\gamma_{\rm max}|>|\sigma|$ & synchr.\ unstable & synchr.\ stable\newline if $|\gamma_{\rm max}| < r_0$ \\
\end{tabular}
\end{ruledtabular}
\end{table}
In networks of type (a) and (b) synchronization on a FP or a PO
(stable within the SM) is always stable. For type (c) this dynamics
may be stable or not depending on the particular network topology
(value of $|\gamma_{\rm max}|$) and the dynamics in the SM (value of
$r_0$). On the other hand chaos synchronization is always unstable in
networks of type (b) and (c) and it may be stable or not in networks
of type (a) again depending on the particular network and the
dynamics.

Note that, in contrast to maps, autonomous flows with a stable PO in
the SM always have $r_0 = |\sigma|$, due to the PO's Goldstone
mode. Thus for this case synchronization will be unstable for type (c)
networks. For type (b) networks the stability of the synchronized
solution in this case undergoes  a destabilizing bifurcation.

We now list some examples for the three types of networks. The
classification follows from the eigenvalue structure of the
corresponding coupling matrices $g_{ij}$. Mean field coupled
systems are of type (a), networks with only inhibitory or only
excitatory connections are (up to the row sum factor) stochastic
matrices and are thus of type (a) or (b). Rings of uni-directionally
coupled elements and two bidirectionally coupled elements are of type
(b) and any network with zero row sum ($\sigma=0$) is of type (b)
(trivial case) or (c) and therefore these systems can never exhibit
chaos synchronization.

Another conclusion we can draw from the structure of the MSF confirms
the conjecture stated in \cite{KIN09}: networks with $\sigma=0$
are of type (c) and thus chaos synchronization is always unstable.

Concerning the impact of noise on the delay-coupled network
\cite{HUN10}, for the case of FPs and POs stable synchronization will
be robust to small noise strength.  On the other hand, for the
chaotic case there may exist another radius $r_b<r_0$, where the first
UPO in the attractor loses its transverse stability and the attractor
undergoes a bubbling bifurcation \cite{OTT94,ASH96a}.
Then any network with $r_b < |\gamma_{\rm max}| < r_0$ will exhibit
bubbling in the presence of small noise (or parameter mismatch), while
any network with $|\gamma_{\rm max}|<r_b$ will show stable
synchronization, even in the presence of small noise.  For large noise
strength the linear theory cannot make predictions.

\textsl{Example} -- As an example we consider a network of optically
coupled semiconductor lasers modeled by dimensionless equations of
Lang-Kobayashi \cite{LAN80b} type
\begin{align}
  \dot E^{l}(t) &= \shalf (1+i\alpha)\,n^l(t)\, E^{l}(t) + \sum\nolimits_{j=1}^N g_{lj}\, E^{j}(t-\tau),\nonumber\\
  T\,\dot n^{l}(t) &= p-n^{l}(t)-\left(1+n^{l}(t)\right) |E^{l}(t)|^2 \label{eq:lk},
\end{align}
where $E^l$ and $n^l$ are the complex electric field amplitude and the
inversion of the $l$-th laser, respectively.  For our example, we
choose the parameters as follows: Ratio between carrier and photon
lifetime $T=200$, injection current $p=10$, $\alpha$-factor $\alpha=4$.  
This results in a relaxation oscillation period
$T_{RO}\approx 28$. Figure \ref{fig:pseudo}b shows the $\lambda_{\rm max}=0$
contour line of the corresponding MSF for networks with $\sigma=0.4$ for
different values of the delay time $\tau$. For $\tau=20$ (order of $T_{RO}$) 
the contour line starts to become circular. For $\tau \gtrsim 3 T_{RO}$ 
the shape of the MSF perfectly resembles our predictions.
In this case we find $r_0<\sigma=0.4$, \ie, the
dynamics is chaotic. For $\tau=1$ and $10$ the limit of large delay is
not satisfied, hence the MSF does not exhibit the rotation
symmetry. Note that for these values of the delay time the dynamics is
a PO and since the system is a flow, the stability boundary reaches
its maximum real value at $re^{i\psi}=|\sigma|$.

\textsl{Conclusion} -- We have shown that the MSF has a simple
universal structure in the limit of large delay: it is rotationally
symmetric around the origin and either positive and constant (if it is
positive at the origin), or monotonically increasing and becoming
positive at a critical radius $r_0$. This structure allows us to
confirm a recent conjecture \cite{KIN09} about synchronizability of
chaotic elements. Furthermore, we classify networks into three types
depending on the magnitude of the maximum transversal eigenvalue of
the coupling matrix in relation to the magnitude of the row
sum. Importantly, this classification allows us to predict the
synchronizability of general networks of identical units with strongly
delayed connections based solely on the modulus of the eigenvalues and
the type of synchronized dynamics.  In many cases this prediction is
possible even without computing the critical radius $r_0$ (as shown in
Table~I). Although our results describe the properties of coupled
systems in the limit of large delay, practically they are expected to
hold when the delay is two or three times larger than the
characteristic timescale of the underlying system without delay. This
is confirmed by our example as well as the results of
Refs.~\cite{FAR82GIA96MEN98WOL06YAN06YAN09}.

This work was supported by DFG (Sfb 555 and Research Center MATHEON
under project D21).

\vfill


\end{document}